\documentclass[a4paper,fleqn,usenatbib]{mnras}
\usepackage{newtxtext,newtxmath}
\usepackage[T1]{fontenc}
\usepackage{ae,aecompl}

\usepackage{graphicx}
\usepackage{amsmath}
\usepackage{amssymb}
\usepackage{footnote}

\def\wk{WASP-167/KELT-13}
\def\wkb{WASP-167b/KELT-13b}
\def\prot{$P_{\rm rot}$}
\def\porb{$P_{\rm orb}$}
\def\teff{$T_{\rm eff}$}

\title[WASP-167b/KELT-13b]{WASP-167b/KELT-13b: Joint discovery of a hot Jupiter transiting a rapidly-rotating F1V star}

\author[L. Temple et al.]{
L.Y. Temple,$^{1}$\thanks{E-mail: l.y.temple@keele.ac.uk}
C. Hellier$^{1}$,
M. D. Albrow$^{2}$,\newauthor
D.R. Anderson$^{1}$,
D. Bayliss$^{3}$,
T. G. Beatty$^{4,5}$,
A. Bieryla$^{6}$,\newauthor
D.J.A. Brown$^{7}$,
P. A. Cargile$^{6}$,
A. Collier Cameron$^{8}$,
K. A. Collins$^{9}$,
K. D. Col\'on$^{10,11}$,\newauthor
I. A. Curtis$^{12}$,
G. D'Ago$^{13,14}$,
L. Delrez$^{15,16}$,
J. Eastman$^{6}$,\newauthor
B. S. Gaudi$^{17}$,
M. Gillon$^{15}$,
J. Gregorio$^{18}$,
D. James$^{19}$,\newauthor
E. Jehin$^{15}$,
M. D. Joner$^{20}$,
J. F. Kielkopf$^{21}$,
R. B. Kuhn$^{22,23}$,\newauthor
J. Labadie-Bartz$^{11}$,
D. W. Latham$^{6}$,
M. Lendl$^{24,25}$, 
M. B. Lund$^{9}$,\newauthor
A. L. Malpas$^{2}$,
P.F.L. Maxted$^{1}$,
G. Myers$^{26}$,
T. E. Oberst$^{27}$,\newauthor
F. Pepe$^{25}$,
J. Pepper$^{11}$,
D. Pollacco$^{7}$,
D. Queloz$^{16}$,\newauthor
J. E. Rodriguez$^{6}$,
D. S\'egransan$^{25}$,
R. J. Siverd$^{28}$,
B. Smalley$^{1}$,
K. G. Stassun$^{9,29}$,\newauthor
D. J. Stevens$^{17}$,
C. Stockdale$^{30}$,
T.G. Tan$^{31}$,
A.H.M.J. Triaud$^{25,32}$,\newauthor
S. Udry$^{25}$,
S. Villanueva Jr.$^{17}$,
R.G. West$^{7}$,
G. Zhou$^{6}$
\\
$^{1}$Astrophysics Group, Keele University, Staffordshire, ST5 5BG, UK\\
$^{2}$Department of Physics and Astronomy, University of Canterbury, Private Bag 4800, Christchurch, New Zealand\\
$^{3}$Observatoire Astronomique de l'Universit\'e de Gen\`eve, 51, Chemin des Maillettes, 1290 Versoix, Switzerland\\
$^{4}$Department of Astronomy \& Astrophysics, The Pennsylvania State University, 525 Davey Lab, University Park, PA 16802, USA\\
$^{5}$Center for Exoplanets and Habitable Worlds, The Pennsylvania State University, 525 Davey Lab, University Park, PA 16802, USA\\
$^{6}$Harvard-Smithsonian Center for Astrophysics, 60 Garden Street, Cambridge, MA 02138, USA\\
$^{7}$Department of Physics, University of Warwick, Gibbet Hill Road, Coventry, 
CV4 7AL, UK\\
$^{8}$SUPA, School of Physics and Astronomy, University of St.\ Andrews, North Haugh,  Fife, KY16 9SS, UK\\
$^{9}$Department of Physics and Astronomy, Vanderbilt University, Nashville, TN 37235, USA\\
$^{10}$NASA Goddard Space Flight Center, Greenbelt, MD 20771, USA\\
$^{11}$Department of Physics, Lehigh University, 16 Memorial Drive East, Bethlehem, PA, 18015, USA\\
$^{12}$ICO, Adelaide, Australia\\
$^{13}$Istituto Internazionale per gli Alti Studi Scientifici (IIASS), Via G. Pellegrino 19, 84019 Vietri sul Mare (SA), Italy\\
$^{14}$INAF-Observatory of Capodimonte, Salita Moiariello, 16, 80131, Naples, Italy\\
$^{15}$Institut d'Astrophysique et de G\'eophysique, Universit\'e de Li\`ege, All\'ee du 6 Ao\^ut, 17, Bat. B5C, Li\`ege 1, Belgium\\
$^{16}$Cavendish Laboratory, J J Thomson Avenue, Cambridge, CB3 0HE, UK\\
$^{17}$Department of Astronomy, The Ohio State University, 140 W. 18th Avenue, Columbus, OH 43210, USA\\
$^{18}$Atalaia Group \& CROW Observatory, Portalegre, Portugal\\
$^{19}$Astronomy Department, University of Washington, Box 351580, Seattle, WA 98195, USA\\
$^{20}$Department of Physics and Astronomy, Brigham Young University, Provo, UT 84602, USA\\
$^{21}$Department of Physics and Astronomy, University of Louisville, Louisville, KY 40292 USA\\
$^{22}$South African Astronomical Observatory, PO Box 9, Observatory, 7935 Cape Town, South Africa\\
$^{23}$Southern African Large Telescope, PO Box 9, Observatory, 7935 Cape Town, South Africa\\
$^{24}$Space Research Institute, Austrian Academy of Sciences, Schmiedlstr. 6, 80 42, Graz, Austria\\
$^{25}$Observatoire astronomique de l'Universit\'e de Gen\`eve 51 ch. des Maillettes, 1290 Sauverny, Switzerland\\
$^{26}$AAVSO, 49 Bay State Rd. Cambridge, MA 02138\\
$^{27}$Department of Physics, Westminster College, New Wilmington, PA 16172\\
$^{28}$Las Cumbres Observatory Global Telescope Network, 6740 Cortona Dr., Ste 102, Goleta, CA 93117, USA\\
$^{29}$Department of Physics, Fisk University, 1000 17th Ave. N., Nashville, TN 37208, USA\\
$^{30}$Hazelwood Observatory, Churchill, Victoria, Australia\\
$^{31}$Perth Exoplanet Survey Telescope, Perth, Australia\\
$^{32}$Institute of Astronomy,  University of Cambridge,  Cambridge, CB3 0HA, UK
}

\date{Accepted XXX. Received YYY; in original form ZZZ}

\pubyear{2017}

\begin{document}
\label{firstpage}
\pagerange{\pageref{firstpage}--\pageref{lastpage}}
\maketitle

\begin{abstract}
We report the joint WASP/KELT discovery of \wkb, a transiting hot Jupiter with a 2.02-d orbit around a $V$\,=\,10.5, F1V star with [Fe/H]\,=\,0.1\,$\pm$\,0.1.  The 1.5\,R$_{\rm Jup}$ planet was confirmed by Doppler tomography of the stellar line profiles during transit.  We place a limit of $<$\,8\,M$_{\rm Jup}$ on its mass.   The planet is in a retrograde orbit with a sky-projected spin--orbit angle of $\lambda\,=\,-165^{\circ}\,\pm\,5^{\circ}$. This is in agreement with the known tendency for orbits around hotter stars to be more likely to be misaligned.  \wk\ is one of the few  systems where the stellar rotation period is less than the planetary orbital period.  We find evidence of non-radial stellar pulsations in the host star, making it a $\delta$-Scuti or $\gamma$-Dor variable.   The similarity to WASP-33, a previously known hot-Jupiter host with pulsations, adds to the suggestion that close-in planets might be able to excite stellar pulsations. 
\end{abstract}


\section{Introduction}
\label{sec:intro}
There are far fewer hot-Jupiter exoplanets known to transit hot stars with \teff\,$>$\,6700\,K than those transiting later-type stars. This is partially a selection effect given that planets transiting very hot or fast-rotating stars are harder to validate, since the lack of spectral lines makes it harder to obtain accurate radial-velocity measurements. Thus, radial-velocity surveys have tended to avoid hotter stars, while transit searches such as the Wide Angle Search for Planets (WASP) have, in the past, paid less attention to such candidates. 

Hot Jupiters are often in orbits that are not aligned with the stellar rotation axis.  One explanation is that hot Jupiters migrate within a disc which is itself tilted with respect to the stellar rotation axis, possibly due to a companion \citep{2014sf2a.conf..217C,2015MNRAS.450.3306F}. Another is that  hot Jupiters arrive at their current orbits through high-eccentricity migration, owing to perturbations by third bodies \citep[e.g.][]{2014ApJ...781L...5D}, which leads to a range of orbital obliquities. 

The planets that have been found around hotter stars have a greater tendency to be in misaligned orbits, compared to planets orbiting later-type stars \citep{2010ApJ...718L.145W,2012ApJ...757...18A}, suggesting a systematic difference in their dynamical history.    In addition there appears to be a dearth of hot Jupiters orbiting very fast rotators, such that the star rotates faster than the planetary orbit \citep[e.g.][]{2003ApJ...589..605W,2007ApJ...669.1298F,2013ApJ...775L..11M}, though this may again be partially a selection effect. In such stars the usual tidal decay of a hot Jupiter orbit would be reversed, provided the orbit is prograde, again changing the dynamical history.  

For such reasons the WASP \citep{2006PASP..118.1407P,2011EPJWC..1101004H} and KELT \citep[Kilodegree Extremely Little Telescope:][]{2007PASP..119..923P, 2012PASP..124..230P} transit-search teams are now giving more attention to hotter candidates.   The first hot Jupiter found to transit an A-type star was WASP-33b, where the planet was validated, not by radial-velocity measurements, but by the detection of the shadow of the planet seen through tomography of the stellar line profiles during transit \citep{2010MNRAS.407..507C}.  This technique requires a higher signal-to-noise ratio than radial-velocity measurements, and thus a bigger telescope for a given host-star magnitude. 

Tomographic methods have since led to the detection of the hot Jupiters KELT-17b \citep{2016arXiv160703512Z}, HAT-P-57b \citep{2015AJ....150..197H}, XO-6b \citep{2016arXiv161202776C}, HAT-P-67b \citep{2017arXiv170200106Z} and most recently KELT-9b \citep{2017Natur.546..514G}, as well as the warm Jupiter Kepler-448b \citep{2015A&A...579A..55B}.   The hot Jupiter Kepler-13 Ab has also been detected tomographically \citep{2014ApJ...790...30J}, though in that case the planet's existence had previously been confirmed using the orbital phase curve \citep{2011AJ....142..195S,2012A&A...541A..56M}.  

In this work we present the joint WASP/KELT  discovery of a transiting hot Jupiter dubbed \wkb .  The planet host star is a 7000 K, rapidly rotating ($v\,\sin i_\star\,\approx$\,50\,km\,s$^{-1}$) F1V star. 

\section{Data and Observations}
\label{sec:data}
WASP-167b/KELT-13b was observed with WASP-South from 2006 May--2012 June and with KELT-South from 2010 March--2013 August. WASP-South is an eight-camera array using 200-mm f/1.8 lenses, covering a $7.8^{\circ}\times7.8^{\circ}$ field of view. Typically eight fields per night were observed with a broad-band filter (400--700nm) using 30-s exposures and typically 10-minute cadence. Details of the data reduction and processing are given by \citet{2006MNRAS.373..799C} and an explanation of the process for selecting candidates is given by \citet{2007MNRAS.380.1230C}.

The KELT-South site consists of a single 80-mm f/1.9 camera with a $26^{\circ}\times26^{\circ}$  field of view and a pixel scale of 23''. Survey observations use 150-s exposures and a cadence of 10--20 minutes per field. Further details of KELT-South are given in \citet{2012PASP..124..230P}. Details of the data reduction, processing and candidate selection procedures are given by \citet{2012ApJ...761..123S} and \citet{2016MNRAS.459.4281K}.

\begin{figure}
\hspace*{2mm}\includegraphics[width=0.48\textwidth]{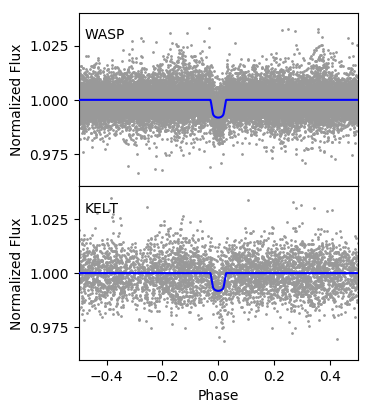}\\ [-2mm]
\caption{The WASP (top) and KELT (bottom) discovery light curves for WASP 167b/KELT-13b, folded on the orbital period. The blue lines show the final model obtained in the MCMC fitting (see Section~\ref{sec:photanalysis}).}
\label{fig:discphot}
\end{figure}

The WASP and KELT teams independently found a planet-like transit signal with a $\sim$\,2-day period (see Fig.~\ref{fig:discphot}) and set about obtaining a total of 18 follow-up light curves of the transit. The observations are listed in Table~\ref{table:observations} while the lightcurves are shown in Fig.~\ref{fig:followupphot}.  The techniques for obtaining relative photometry have been reported in previous WASP and KELT discovery papers, and since we have 18 transit curves from disparate facilities we refer the reader to such papers for full details of the instrumentation and analysis \citep[e.g.][]{2014MNRAS.440.1982H,2016A&A...591A..55M, 2016MNRAS.459.4281K,2016arXiv160701755P,2016AJ....151..138R}.  We give key details of the instrumentation used in Table~\ref{table:observations}.

In an attempt to refute the planetary hypothesis we, on three occasions, attempted to detect an eclipse (of the occulting body by the star) using TRAPPIST with a $z^{\prime}$ filter (see Table ~\ref{table:observations} for details).  This is discussed in Sec.~\ref{sec:photanalysis}. 

The two teams also began monitoring the radial velocity of the star using the Euler/CORALIE and TRES spectrographs \citep{2001Msngr.105....1Q, Furesz2008}. The measured values are listed in Table~\ref{table:RVs}. The crucial tomographic data, revealing the planet shadow, then came from an observation over a transit on the night of March 1$^{\rm st}$ 2016 using the ESO 3.6-m/HARPS spectrograph \citep{2002Msngr.110....9P}.

We have searched the combined WASP and KELT photometry of \wk\ for modulations indicating the rotational period of the star, as described by \citet{2011PASP..123..547M}, but did not find any modulations above $\sim$\,0.7\,mmag at periods longer than 1 day.

\begin{table*}
\caption{Details of all observations of WASP-167b/KELT-13b used in this work, including the discovery photometry, the follow-up photometry and the spectroscopic observations. The label in the final column corresponds to a lightcurve in Fig.~\ref{fig:followupphot}.}
\centering
\begin{tabular}{lccccccc}
\hline\hline
Facility & Location & Aperture & FOV & Pixel Scale & Date & Notes & Label \\
 & & & ($^{\prime} \times ^{\prime}$) & ( $^{''}$pixel$^{-1}$) & & & \\ [0.5ex]
\hline
\multicolumn{8}{l}{{\it Discovery Photometry}} \\
WASP-South & SAAO$^{1}$, South Africa & 111\,mm & $7.8\times7.8$ & 14 & 2006 May-- & 26114 points & - \\
 & & & & & 2012 Jun & & \\
KELT-South & SAAO, South Africa & 42\,mm & $26\times26$ & 23 & 2010 Mar-- & 4563 points & - \\
 & & & & & 2013 Aug & & \\
\hline
\multicolumn{1}{l}{{\it Transit observations}} \\
TRAPPIST & ESO$^{2}$, La Silla, Chile & 0.6\,m & $22\times22$ & 0.65 & 2012 Feb 22 & I+z' & a \\
TRAPPIST & ESO, La Silla, Chile & 0.6\,m & $22\times22$ & 0.65 & 2012 Apr 30 & I+z' & b \\
LCOGT-LSC & CTIO$^{3}$, Chile & 1\,m & $26.5\times26.5$ & 0.4 & 2014 May 17 & i' & c \\
PEST & Perth, Australia & 0.3\,m & $31\times21$ & 1.2 & 2014 Jun 22 & Rc & d \\
PEST & Perth, Australia & 0.3\,m & $31\times21$ & 1.2 & 2015 Jan 14 & V & e \\
Skynet/Prompt4 & CTIO, Chile & 0.4\,m & $10\times10$ & 0.59 & 2015 Feb 22 & z' & f \\
T50 Telescope & SSO$^{4}$, Australia & 0.43\,m & $16.2\times15.7$ & 0.92 & 2015 Mar 24 & B & g \\
T50 Telescope & SSO, Australia & 0.43\,m & $16.2\times15.7$ & 0.92 & 2015 Mar 26 & B & h \\
Mt. John & UC$^{5}$, New Zealand & 0.6\,m & $14\times14$ & 0.549 & 2015 Mar 26 & V & i \\
LCOGT-COJ & SSO, Australia & 1\,m & $15.8\times15.8$ & 0.24 & 2015 Mar 28 & r' & j \\
PEST & Perth, Australia & 0.3\,m & $31\times21$ & 1.2 & 2015 Mar 28 & Ic & k \\
LCOGT-COJ & SSO, Australia & 1\,m & $15.8\times15.8$ & 0.24 & 2015 Mar 28 & i' & l \\
Hazelwood & Victoria, Australia & 0.32\,m & $18\times12$ & 0.73 & 2015 Mar 30 & B & m \\
Ivan Curtis & Adelaide, Australia & 0.235\,m & $16.6\times12.3$ & 0.62 & 2015 Mar 30 & V & n \\
Ellinbank & Victoria, Australia & 0.32\,m & $30.4\times14.1$ & 1.12 & 2015 Apr 03 & B & o \\
PEST & Perth, Australia & 0.3\,m & $31\times21$ & 1.2 & 2015 Apr 03 & B & p \\
LCOGT-CPT & SAAO, South Africa & 1\,m & $15.8\times15.8$ & 0.24 & 2015 Apr 17 & Z & q \\
TRAPPIST & ESO, La Silla, Chile & 0.6\,m & $22\times22$ & 0.65 & 2016 Mar 01 & z' & r \\
\hline
\multicolumn{8}{l}{{\it Occultation window observations}} \\
TRAPPIST & ESO, La Silla, Chile & 0.6\,m & $22\times22$ & 0.65 & 2011 Feb 13 & z' & - \\
TRAPPIST & ESO, La Silla, Chile & 0.6\,m & $22\times22$ & 0.65 & 2011 Apr 25 & z' & - \\
TRAPPIST & ESO, La Silla, Chile & 0.6\,m & $22\times22$ & 0.65 & 2011 May 09 & z' & - \\
\hline
\multicolumn{8}{l}{{\it Spectroscopic Observations}} \\
CORALIE	& ESO, La Silla, Chile & 1.2\,m & - & - & 2010 Apr-- & 21 RVs & - \\
 & & & & & 2017 Mar & & \\
TRES & FLWO$^{6}$, Arizona & 1.5\,m  & - & - & 2015 Feb-- & 20 RVs & - \\
 & & & & & 2016 Apr & & \\
HARPS & ESO, La Silla, Chile & 3.6\,m & - & - & 2016 Mar 01 & 17 CCFs & - \\ [1ex]
\hline
\end{tabular}
$^{1}$South African Astronomical Observatory, $^{2}$European Southern Observatory, $^{3}$Cerro Tololo Inter-American Observatory, $^{4}$Siding Spring Observatory, $^{5}$University of Canturbury, $^{6}$Fred Lawrence Whipple Observatory
\label{table:observations}
\end{table*}

\begin{table}
\caption{Radial velocities and bisector spans for \wkb\ .}
\centering
\begin{tabular}{lcccc}
\hline\hline
BJD & RV & $\sigma$$_{\rm RV}$ & BS & $\sigma$$_{\rm BS}$ \\
(TDB)  & (km\,s$^{-1}$) & (km\,s$^{-1}$) & (km\,s$^{-1}$) & (km\,s$^{-1}$) \\ [0.5mm]
\hline
\multicolumn{5}{l}{TRES RVs:} \\
2457055.9950 & --1.01 & 0.41 & --0.12 & 0.33 \\
2457057.0280 & 0.00$^*$ & 0.23 & 0.03 & 0.21 \\
2457058.0115 & --0.57 & 0.28 & 0.53 & 0.37 \\
2457060.9845 & --0.39 & 0.31 & 0.33 & 0.21 \\
2457086.9123 & --0.63 & 0.29 & 0.04 & 0.24 \\
2457122.8138 & --1.85 & 0.42 & --0.05 & 0.17 \\
2457123.8544 & --1.15 & 0.39 & --0.23 & 0.28 \\
2457137.7848 & --2.39 & 0.24 & --0.29 & 0.14 \\
2457139.7803 & --2.00 & 0.39 & 0.15 & 0.24 \\
2457141.7805 & --1.27 & 0.45 & 0.17 & 0.11 \\
2457143.7671 & --2.19 & 0.23 & 0.17 & 0.14 \\
2457144.7561 & --1.81 & 0.35 & --0.05 & 0.18 \\
2457145.7548 & --1.39 & 0.32 & --0.02 & 0.16 \\
2457149.7475 & --1.16 & 0.42 & 0.12 & 0.25 \\
2457150.7423 & --2.33 & 0.44 & 0.05 & 0.20 \\
2457151.7486 & --1.19 & 0.44 & --0.20 & 0.23 \\
2457152.7481 & --2.52 & 0.36 & --0.29 & 0.18 \\
2457406.0418 & --1.03 & 0.34 & -- & -- \\
2457491.8087 & --0.84 & 0.34 & -- & -- \\
2457504.7985 \medskip & --1.62 & 0.28 & - & - \\
\hline
\multicolumn{5}{l}{CORALIE RVs:} \\
2455310.5205 & --3.82 & 0.059 & --2.38 & 0.12 \\
2455310.8005 & --3.74 & 0.061 & --0.42 & 0.12 \\
2455311.8282 & --2.83 & 0.059 & --1.30 & 0.12 \\
2455320.5338 & --3.50 & 0.063 & 1.80 & 0.13 \\
2455320.7628 & --2.23 & 0.066 & --2.30 & 0.13 \\
2455568.8079 & --2.72 & 0.060 & --2.70 & 0.12 \\
2455572.8753 & --3.09 & 0.066 & -- & -- \\
2455574.8568 & --3.57 & 0.061 & --4.96 & 0.12 \\
2455646.7729 & --2.51 & 0.070 & --2.96 & 0.14 \\
2455712.5532 & --3.36 & 0.059 & 0.26 & 0.12 \\
2455722.5360 & --2.27 & 0.058 & --4.94 & 0.12 \\
2455979.6816 & --3.76 & 0.062 & --3.44 & 0.12 \\
2455979.8955 & --3.74 & 0.059 & --7.50 & 0.12 \\
2455981.7270 & --4.62 & 0.065 & --1.09 & 0.13 \\
2457600.5011 & --4.28 & 0.071 & -- & -- \\
2457616.4971 & --4.70 & 0.066 & 0.048 & 0.13 \\
2457759.8370 & --4.71 & 0.065 & --1.86 & 0.13 \\
2457760.8366 & --3.77 & 0.066 & --3.70 & 0.13 \\
2457804.7057 & --4.08 & 0.065 & -- & -- \\
2457809.7750 & --5.18 & 0.064 & --2.88 & 0.13 \\
2457818.6607 \medskip & --4.73 & 0.067 & --0.25 & 0.13 \\
\hline
\end{tabular}
$^*$ This observation was used as the template for the extraction of the TRES radial velocities.
\label{table:RVs}
\end{table}

\section{Spectral analysis}
\label{sec:specanalysis}
To determine the spectral parameters of the host star we  produced a median-stacked spectrum from the 17 HARPS spectra and used it to find the stellar effective temperature \teff , the stellar metallicity $\rm [Fe/H]$, and the projected stellar rotational velocity $ v \sin i_\star$. The spectra were line-poor and broad-lined, owing to the host star's spectral type, which meant that a  determination of the stellar surface gravity $\log{g_\star}$ was not possible. We therefore assume here a value of $\log{g_\star}\,=\,4.3$, the expected value for a similar star at zero age \citep{1992oasp.book.....G}. The \teff\ was measured using the H-alpha line, which was strong and unblended. The values obtained for each of these parameters are given in Table~\ref{table:results}. Fuller details of our spectral analysis procedure can be found in \citet{2013MNRAS.428.3164D}. We also used the MKCLASS program \citep{2014AJ....147...80G} to obtain a spectral type of F1V.

\section{Photometric and Radial Velocity Analysis}
\label{sec:photanalysis}
We carried out a Markov Chain Monte Carlo (MCMC) fitting procedure, simultaneously modelling the WASP and KELT light curves, the 18 follow-up light curves, and the out-of-transit RVs. We use the latest version of the code described by \citet{2007MNRAS.380.1230C} and \citet{2008MNRAS.385.1576P}.

\begin{figure*}
\hspace*{2mm}\includegraphics[width=0.98\textwidth]{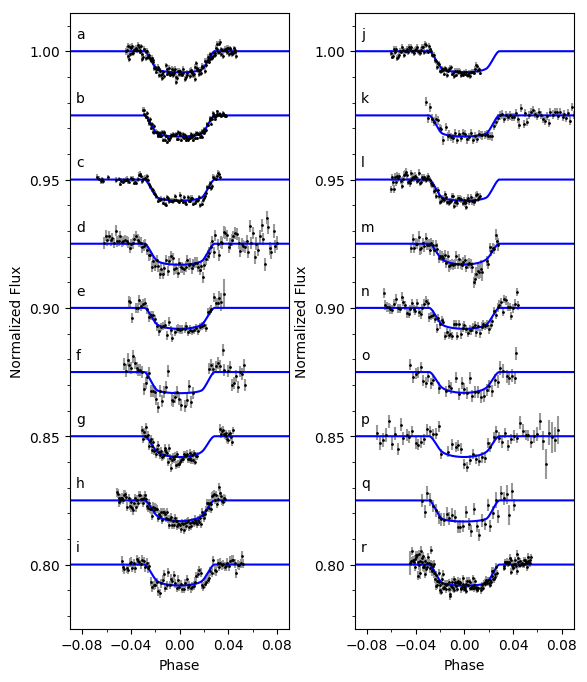}\\ [-2mm]
\caption{The 18 follow-up transit light curves. The blue lines show the final model obtained in the MCMC fitting (see Section~\ref{sec:photanalysis}). The label to the left of each data set corresponds to an entry in the final column of Table~\ref{table:observations}.}
\label{fig:followupphot}
\end{figure*}

Prior to the fit, the KELT team's follow-up lightcurves were detrended by first fitting them using the online EXOFAST applet \citep{2013PASP..125...83E}, removing the effects of airmass and some systematics (this was not needed for TRAPPIST lightcurves). For consistency, we also converted all datasets to the BJD\_TDB time standard using the \citet{2010PASP..122..935E} BJD conversion code. Limb darkening was accounted for using the \citet{2000A&A...363.1081C, 2004A&A...428.1001C} four-parameter non-linear law. At each step of the MCMC the limb-darkening coefficients were interpolated from the Claret tables appropriate to the passband used and the new values of \teff, $\rm [Fe/H]$ and $\log {g_{\star}}$. 

Hot Jupiters settle into a circular orbit on time-scales that are often shorter than their host stars' lifetimes through tidal circularization \citep{2011MNRAS.414.1278P}. We therefore assume a circular orbit, since this will give the most likely parameters \citep{2012MNRAS.422.1988A}. 

The system parameters which determine the shape of the transit light curve are:  the epoch of mid-transit $T_{\rm c}$, the orbital period $P$, the planet-to-star area ratio $(R_{\rm p}/R_{\star})^{2}$ or transit depth $\delta$, the transit duration $T_{\rm 14}$, and the impact parameter $b$. In the RV modelling, we fit the value of the stellar reflex velocity semi-amplitude $K_{\rm 1}$ and the barycentric system velocity $\gamma$. The proposed values of stellar and planetary masses and radii are constrained by the Enoch--Torres relation \citep{2010A&A...516A..33E, 2010A&ARv..18...67T}. We allow for a possible offset in RVs between the CORALIE and TRES datasets. 

Since we collect data from many sources with differing data qualities, our code includes a provision for re-scaling the error bars of each dataset to give $\chi^{2}_{\nu}$ = 1.  This means that datasets that don't fit as well are down-weighted, such that the final result is dominated by the better datasets. With 18 transit lightcurves, this means that the final parameters are relatively insensitive to red noise in particular lightcurves.   

\begin{figure}
\hspace*{2mm}\includegraphics[width=0.49\textwidth]{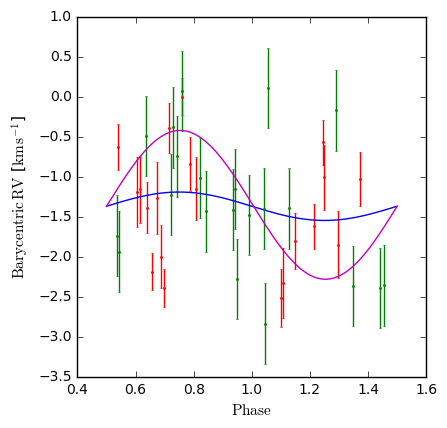}\\ [-2mm]
\caption{The 21 CORALIE RVs (green) and 20 TRES RVs (red) obtained for WASP-167/KELT-13. The blue line shows the best fitting semi-amplitude, which we do not regard as reliable.  The magenta line shows the RV amplitude for a planet mass of 8\,$M_{\rm Jup}$, which we regard as a conservative upper limit.}
\label{fig:RV-maxK1}
\end{figure}

The radial velocities and the best-fitting model are shown in Figure~\ref{fig:RV-maxK1}. There is a clear scatter in the RVs about the model, beyond that attributable to the error bars. This could, for example, be caused by the pulsations in the host star distorting the stellar line profiles (see Sections~\ref{sec:dopanalysis} and~\ref{discuss:pulsations}), or by a third body in the system. 

Attempting to fit for a second planet does not properly explain the scatter, but does significantly change the semi-amplitude fitted to the first planet.  For this reason we do not regard the fitted semi-amplitude as a reliable measure of the planet's mass, but instead report an upper limit of 8\,M$_{\rm Jup}$, which we regard as conservative but sufficient to demonstrate that the transiting body has a planetary mass.   We are continuing to monitor the system in order to discover the cause of the scatter.   The parameters obtained in this analysis are given in Table~\ref{table:results}. 

The TRAPPIST observations of the eclipse (of the planet by the star) produced no detection, with an upper limit of 1100\,ppm.   Given the stellar and planetary radii (Table~\ref{table:results}) this implies that the heated face of the planet must be cooler than 3750\,K.  The fitted system parameters imply a planet temperature of T$_{\rm eql}$ = 2330\,$\pm$\,65\,K, and thus the non-detection of the eclipse is consistent with the planetary hypothesis.

\section{Doppler Tomography}
\label{sec:dopanalysis}
We obtained 17 spectra with the ESO 3.6-m/HARPS spectrograph through a transit on the night of March 1$^{\rm st}$ 2016. We also observed the same transit photometrically using TRAPPIST (see lightcurve r in Fig.~\ref{fig:followupphot}, Table ~\ref{table:observations}). The standard HARPS  Data Reduction Software produces a  cross correlation function (CCF)  correlated over a window of $\pm$\,300\,km\,s$^{-1}$ (as described in \citet{1996A&AS..119..373B}, \citet{2002Msngr.110....9P}). The CCFs were created using a mask matching a G2 spectral type, containing zeroes at the positions of absorption lines and ones in the continuum.

\begin{figure}
\centering
\hspace*{2mm}\includegraphics[width=0.49\textwidth]{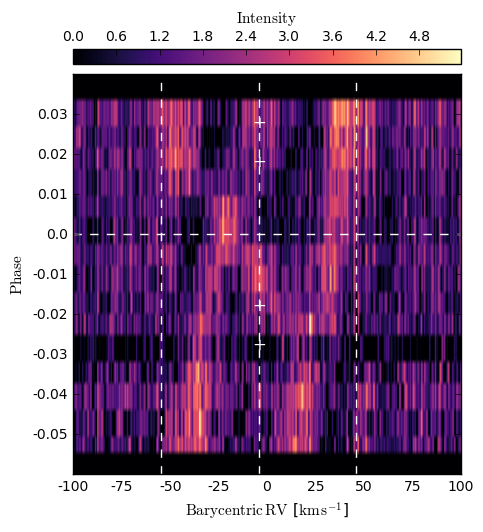}\\ [-2mm]
\caption{The line profiles through transit. We interpret this as showing prograde moving stellar pulsations and a retrograde moving planet trace. The white dashed vertical lines mark the positions of the $\gamma$ velocity of the system and the positions of $\gamma\,\pm\,v \sin i_{\star}$ (i.e. the centre and edges of the stellar line profile). The phase of mid-transit is marked by the white horizontal dashed line. The white + symbols indicate the four transit contact points, calculated using the ephemeris obtained in the analysis in Section~\ref{sec:photanalysis}.}
\label{fig:SW1304tomog1}
\end{figure}

We display the resulting CCFs as a function of the planet's orbital phase in Fig.~\ref{fig:SW1304tomog1}, where phase 0 is mid-transit. In producing this plot we have first subtracted the invariant part of the CCF profile.  We do this by constructing a ``minimum CCF'', which at each wavelength has the lowest value from the range of phases. 

We interpret the CCFs as showing stellar pulsations moving in a prograde direction (moving redward over time).  Similar pulsations are seen in the tomograms of WASP-33 \citep{2010MNRAS.407..507C,2015ApJ...810L..23J}, which is regarded as a $\delta$-Scuti pulsator \citep{2011A&A...526L..10H}.

To try to remove the pulsations by separating the features into prograde-moving and retrograde components we followed the method of \citet{2015ApJ...810L..23J}, adopted for WASP-33, by Fourier transforming the CCFs, such that the prograde and retrograde components appear in different quadrants in velocity space. 

This separation technique will not be perfect, and we expect some residual contamination from the pulsations.  We thus experimented with which data to include. We found that we get the best separation of the components and thus the clearest planetary signal if we do not include in the Fourier transform the last two spectra.  These were in any case obtained outside the transit and so cannot contain information about a planet. It is thus valid to try Fourier transforms both with and without these two, in order to see which better separates the pulsations and leaves the clearest planet trace.

\begin{figure}
\centering
\hspace*{2mm}\includegraphics[width=0.49\textwidth]{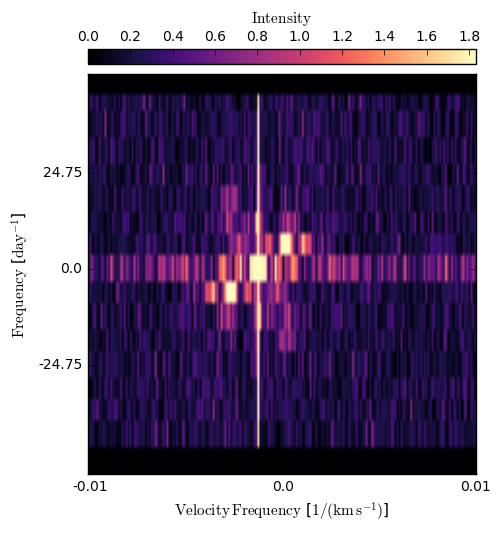}\\ [-2mm]
\caption{The Fourier transform of the line profiles. The stellar pulsations are seen as the diagonal feature from bottom-left to top-right.  The weaker diagonal feature running bottom-right to top-left is produced by the planet. }
\label{fig:SW1304tomog2}
\end{figure}

Fig.~\ref{fig:SW1304tomog2} shows the Fourier-transformed data, where the feature running from bottom-left to top-right can be attributed to the pulsations. We thus applied the filter used by \citet{2015ApJ...810L..23J}, which contained zeroes in the quadrants containing the prograde pulsation signal and unity in the quadrants containing the retrograde signal, with a Hann function bridging the discontinuity.

\begin{figure}
\centering
\hspace*{2mm}\includegraphics[width=0.49\textwidth]{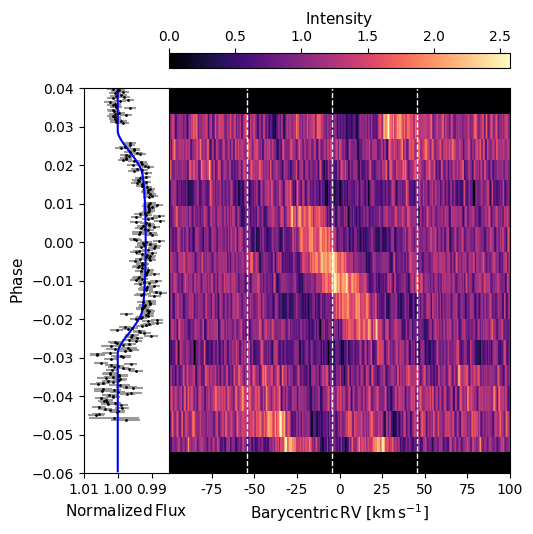}\\ [-2mm]
\caption{The spectral profiles through transit after removing the stellar pulsations via Fourier filtering.   The planet trace is then readily seen moving in a retrograde direction.  The left-hand panel shows the simultaneous TRAPPIST photometry of the transit. }
\label{fig:SW1304tomog3}
\end{figure}

We then Fourier transform the masked data back into phase versus velocity and display that in Fig.~\ref{fig:SW1304tomog3}. This shows an apparent retrograde trace, which we attribute to the shadow of a planet. This is again similar to what is seen in WASP-33 \citep{2015ApJ...810L..23J}.

The planet's Doppler shadow seems to disappear towards the end of the transit (see Fig.~\ref{fig:SW1304tomog3}).  It is likely that it has been reduced during the filtering process, as a result of imperfect separation of the planetary and stellar-pulsation signals. This might have some effect on fitting the alignment angle $\lambda$, which depends on the slope of the Doppler shadow, but should have less effect on the other fitted quantities. 

In order to parametrise the planet's orbit we then fitted the CCFs through transit, in a manner similar to the methods in \citet{2017MNRAS.464..810B}. Since we had subtracted the ``minimum CCF'' above, we first add that back in to the filtered CCFs in order to reintroduce the stellar line profile, which is a key feature for constraining the value of $v \sin i_{\star}$. 

The parameters which define the shape of the CCF line profile are: the projected spin-orbit misalignment angle $\lambda$; the stellar line-profile Full-Width at Half-Maximum (FWHM); the FWHM of the line perturbation due to the planet $v_{\rm FWHM}$; the stellar $\gamma$-velocity, and $v \sin i_\star$. These parameters were fitted using a MCMC fitting algorithm which assumes a Gaussian shape for the line perturbation caused by the planet. Both $v \sin i_\star$ and $v_{\rm FWHM}$ have two data constraints, one from the shape of the line-broadening profile, and one from the slope of the trajectory of the bump across the line profile (given knowledge of $\lambda$). The value of $v \sin i_\star$ obtained in the spectral analysis was used as a prior in the fit. Initial values for the stellar line FWHM and the $\gamma$-velocity were obtained by fitting a Gaussian profile to the CCFs. The $\lambda$ angle and $v_{\rm FWHM}$ were given no prior. Details of the fitting algorithm are given in \citet{2010MNRAS.403..151C}, and the resulting system parameters are listed in Table~\ref{table:results}. 

\begin{figure}
\centering
\hspace*{2mm}\includegraphics[width=0.49\textwidth]{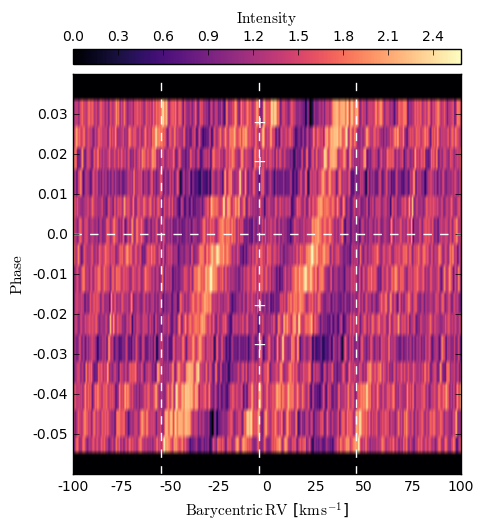}\\ [-2mm]
\caption{The spectral profiles through transit after removing the planet shadow via Fourier filtering.   The stellar pulsations are seen moving in a prograde direction.}
\label{fig:SW1304tomog4}
\end{figure}

Lastly, in Fig.~\ref{fig:SW1304tomog4} we show the pulsations without the planet trace, obtained by filtering to leave only the prograde quadrants, and then transforming back into velocity space. 

\begin{table} 
\caption{System parameters obtained for WASP-167b/KELT-13b in this work.} 
\label{table:results}
\begin{tabular}{lc}
\multicolumn{2}{l}{1SWASP\,J130410.53--353258.2}\\
\multicolumn{2}{l}{2MASS\,J13041053--3532582}\\
\multicolumn{2}{l}{RA\,=\,13$^{\rm h}$04$^{\rm m}$10.53$^{\rm s}$, 
Dec\,=\,--35$^{\circ}$32$^{'}$58.28$^{''}$ (J2000)}\\
\multicolumn{2}{l}{$V$ = 10.5}  \\
\multicolumn{2}{l}{IRFM \teff\ = 6998 $\pm$ 151 K}  \\ 
\multicolumn{2}{l}{{\it Gaia} Proper Motions: (RA) --19.0\,$\pm$\,1.4 mas}\\
\multicolumn{2}{l}{Dec) 0.66\,$\pm$\,1.24 mas/yr}\\
\multicolumn{2}{l}{Parallax: 2.28\,$\pm$\,0.62 mas}\\
\multicolumn{2}{l}{Rotational Modulations: < 0.7 mmag (95\%)}\\
\hline\hline
Parameter (Unit) & Value \\ 
\hline 
\\
\multicolumn{2}{l}{Stellar parameters from spectral analysis:} \\[0.5ex]
\teff\ (K) & 6900\,$\pm$\,150 \\
$\log {A(\rm Fe)}$ & 7.46\,$\pm$\,0.18 \\
{[Fe/H]} & --0.04\,$\pm$\,0.18 \\
$v \sin i_{\rm *}$ (km\,s$^{-1}$) \medskip & 52\,$\pm$\,8 \\
\multicolumn{2}{l}{Parameters from photometry and RV analysis:} \\[0.5ex]
$P$ (d) & 2.0219596\,$\pm$\,0.0000006 \\
$T_{\rm c}$ (BJD) & 2456592.4643\,$\pm$\,0.0002 \\
$T_{\rm 14}$ (d) & 0.1135\,$\pm$\,0.0008 \\
$T_{\rm 12}=T_{\rm 34}$ (d) & 0.0212\,$\pm$\,0.0010 \\
$\Delta\,F=R_{\rm P}^{2}$/R$_{*}^{2}$ & 0.0082\,$\pm$\,0.0001 \\
$b$ & 0.77\,$\pm$\,0.01 \\
$a$ (AU)  & 0.0365\,$\pm$\,0.0006 \\
$i$ ($^\circ$) & 79.9\,$\pm$\,0.3 \\
\teff\ (K) & 7000\,$\pm$\,250 \\
$\log g_{\rm *}$ (cgs) & 4.13\,$\pm$\,0.02 \\
$\rho_{\rm *}$ ($\rho_{\rm \odot}$) & 0.28\,$\pm$\,0.02 \\
{[Fe/H]} & 0.1\,$\pm$\,0.1 \\
$M_{\rm *}$ ($M_{\rm \odot}$) & 1.59\,$\pm$\,0.08 \\
$R_{\rm *}$ ($R_{\rm \odot}$) & 1.79\,$\pm$\,0.05 \\
$T_{\rm eql}$ (K) & 2329\,$\pm$\,64 \\
$M_{\rm P}$ ($M_{\rm Jup}$) & <8 \\
$R_{\rm P}$ ($R_{\rm Jup}$) \medskip & 1.58\,$\pm$\,0.05 \\
\multicolumn{1}{l}{Parameters from tomography:} \\[0.5ex]
$\gamma$ (km\,s$^{-1}$) & --3.409\,$\pm$\,0.007 \\
$v \sin i_{\rm *}$ (km\,s$^{-1}$) & 49.94\,$\pm$\,0.04 \\
$\lambda$ ($^\circ$) & --165\,$\pm$\,5 \\
$v_{\rm FWHM}$ (km\,s$^{-1}$) & 20.9\,$\pm$\,0.9 \\
\\
\hline 
\end{tabular} 
\end{table} 

\section{Evolutionary status}
\label{sec:sedanalysis}
We then used {\sc MINES}{\lowercase {\it weeper}}, a newly developed Bayesian approach to determining stellar parameters using the MIST stellar evolution models \citep{2016ApJ...823..102C}. Examples of the use of {\sc MINES}{\lowercase {\it weeper}} in determining stellar parameters are shown in \citet{2017arXiv170103807R,2017ApJ...836..209R}. We model the available B$_{T}$, V$_{T}$ photometry from Tycho-2, J, H, K$_{s}$ from 2MASS, and WISE W1-3 photometry. We also include in the likelihood calculation the measured parameters from the spectroscopic analysis (\teff\,=\,6900\,$\pm$\,150\,K and $\rm [Fe/H]$\,=\,-0.04$\pm$\,0.18), as well as the {\it Gaia} DR1 parallax ($\pi$\,=\,2.28\,$\pm$\,0.62\,$\rm mas$;  \citet{2016A&A...595A...2G,2016A&A...595A...1G}) and the fitted transit stellar density (0.28\,$\pm$\,0.02\,$\rho_{\rm \odot}$). We applied non-informative priors on all parameters within the MIST grid of stellar evolution models, and a non-informative prior on extinction (A$_{V}$) between 0--2.0\,mags. Our final parameters are determined from the value at the highest posterior probability for each parameter, and the errors are based on the marginalized inner-68th percentile range. These are given in Table ~\ref{table:SED}.

\begin{figure}
\centering
\hspace*{2mm}\includegraphics[width=0.49\textwidth]{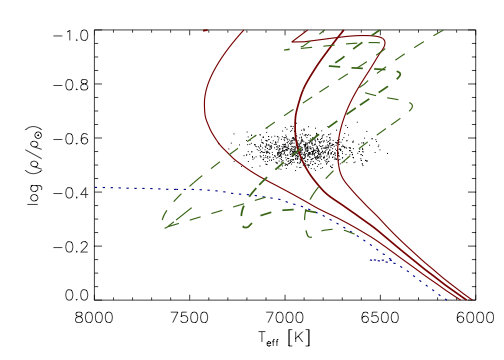}\\ [-2mm]
\caption{The best fitting evolutionary tracks and isochrones of \wk\ obtained using {\sc bagemass}. Dotted blue line: ZAMS at best-fit [Fe/H]. Green dashed lines: evolutionary track for the best-fit [Fe/H] and mass, plus $1\sigma$ bounds. Red lines: isochrone for the best-fit [Fe/H] and age, plus $1\sigma$ bounds.}
\label{fig:trho_plot}
\end{figure}

For comparison, we also use the open source software {\sc bagemass}\footnote{\url{http://sourceforge.net/projects/bagemass}}, which uses the Bayesian method described by \citet{2015A&A...575A..36M}, to estimate the stellar age and mass. The models used in {\sc bagemass} were calculated using the {\sc garstec} stellar evolution code \citep{2008Ap&SS.316...99W}. We use the grid of stellar models in {\sc bagemass} and use the same temperature, metallicity and density constraints as for the MINESweeper calculation. We also apply a luminosity constraint of $\log L_{\rm T}$\,=\,1.00$^{+0.28}_{-0.22}$, which was derived using the {\it Gaia} parallax and the total line-of-sight reddening as determined by \citet{2011ApJ...737..103S,2014MNRAS.437.1681M} (E(B--V)\,=\,0.051\,$\pm$\,0.034). The resulting age and mass values are in Table ~\ref{table:SED}. Both values are compatible with those from MINESweeper. The best-fit stellar evolution tracks and isochrones are shown in Fig.~\ref{fig:trho_plot}.

\begin{table} 
\caption{Stellar parameters obtained for \wk\ in the SED analysis (see Section ~\ref{sec:sedanalysis}).} 
\centering
\label{table:SED}
\begin{tabular}{lc}
\hline\hline
Parameter (Unit) & Value \\ 
\hline
\\
\multicolumn{2}{l}{{\sc MINES}{\lowercase {\it weeper}}:} \\[0.5ex]
Age (Gyr) & 1.29$^{+0.36}_{-0.27}$ \\[0.5ex]
$M_{\rm *}$ ($M_{\rm \odot}$) & 1.518$^{+0.069}_{-0.087}$  \\[0.5ex]
$R_{\rm *}$ ($R_{\rm \odot}$) & 1.756$^{+0.067}_{0.057}$ \\[0.5ex]
$\log L_{\rm *}$ ($L_{\rm \odot}$) & 0.835$^{+0.040}_{-0.034}$  \\[0.5ex]
\teff (K) & 7043$^{+89}_{-68}$ \\[0.5ex]
$\log g_{\rm *}$ & 4.131$^{+0.018}_{-0.028}$ \\[0.5ex]
$\rm{[Fe/H]}_{surface}$ & --0.01$^{+0.17}_{-0.10}$ \\[0.5ex]
$\rm{[Fe/H]}_{init}$ & --0.04$^{+0.16}_{-0.09}$ \\[0.5ex]
Distance (pc) & 381$^{+15}_{-13}$ \\[0.5ex]
A$_{V}$ (mag) & 0.044$^{+0.057}_{-0.025}$ \\
\\
\multicolumn{2}{l}{{\sc bagemass}:} \\[0.5ex]
Age (Gyr) & 1.56\,$\pm$\,0.40 \\[0.5ex]
$M_{*}$ ($M_{\rm \odot}$) & 1.49\,$\pm$\,0.09 \\
\\
\hline
\end{tabular} 
\end{table}

\section{Discussion}
\subsection{Stellar rotation rate and tidal interaction}
As an F1V star with \teff\,=\,6900\,$\pm$\,150\,K, \wk\ is among the hottest stars known to host a transiting hot Jupiter.  Others include WASP-33 \citep{2010MNRAS.407..507C}, Kepler-13 \citep{2011AJ....142..195S,2014ApJ...788...92S}, KELT-17 \citep{2016arXiv160703512Z}, HAT-P-57 \citep{2015AJ....150..197H} and KELT-9 \citep{2017Natur.546..514G}.

In addition, \wk\ appears to be one of the most rapidly rotating stars known to host a hot Jupiter, and one of the few with a stellar rotation period shorter than the planet's orbit.  The measured $v \sin i_\star$ of 49.94\,$\pm$\,0.04\,km\,s$^{-1}$ and the fitted radius of 1.79\,$\pm$\,0.05\,R$_{\odot}$ imply a rotation period of \prot\,<\,1.81-d, which compares with the planet's orbital period of 2.02-d. 

Thus \wk\ joins WASP-33 (\porb\,=\,1.22-d; \prot\,<\,0.79-d, \citealt{2010MNRAS.407..507C}),   KELT-7 (\porb\,=\,2.7-d, \prot\,<\,1.32-d, \citealt{2015AJ....150...12B}) and CoRoT-11b (\porb\,=\,3.0-d; \prot\,<\,1.73-d, \citealt{2010A&A...524A..55G}) in having a hot Jupiter in a $<$\,3-d orbit and an even shorter rotation rate.     See, also, \citet{2016arXiv161202776C} for a discussion of other systems with \prot\,$<$\,\porb\ but with longer period orbits.

The tidal interaction will be different in such systems compared to the more-usual \prot\,$>$\,\porb . In most hot Jupiters, the tidal interaction is expected to drain angular momentum from the orbit, leading to tidal decay of the orbital period \citep[e.g.][]{2009ApJ...692L...9L}. This would be reversed, however, for systems with \prot\,$<$\,\porb, and with the planet in a prograde orbit (such as KELT-7b and CoRoT-11b), thus leading to a different dynamical history.

If, though, \prot\,$<$\,\porb\, and with the planet in a retrograde orbit, such as \wkb\ or WASP-33b, tidal infall would again be expected.    \citet{2013ApJ...775L..11M} analysed {\it Kepler\/} detections and found a dearth of close-in planets around fast rotators, saying that only stars with rotation periods longer than 5--10 days have planets with periods shorter than 3 days.  \citet{2014ApJ...786..139T} then attributed this to the destruction of close-in planets, with the result of spinning up the star.  While \wk\ and the others just named are examples of systems with \prot\,$<$\,\porb\, they are undoubtedly  rare and their dynamics deserves further investigation.

\subsection{The retrograde orbit}
\label{sec:discussLambda}
The planet \wkb\ has a radius of 1.6\,R$_{\rm Jup}$ and is thus inflated, though not exceptionally so.   This is in line with WASP-33b, which has a 1.5\,R$_{\rm Jup}$ radius, and is expected for a hot Jupiter orbiting a hot star, given that a relation between inflated radii and stellar irradiation is now well established \citep[e.g.][]{2011ApJS..197...12D,2012A&A...540A..99E,2016AJ....152..182H}.    We should, though, warn of a selection effect against observing non-inflated planets around relatively large A/F stars, in that the transits would be shallower and may escape detection in WASP-like surveys. 

\citet{2016arXiv161202776C} list 6 planets with measured sky-projected obliquity angles ($\lambda$) that transit host stars hotter than 6700 K (these are XO-6b, CoRoT-3b, KELT-7b, KOI-12b, WASP-33b \&\ Kepler-13Ab).  Of these, five seem to be misaligned but only  moderately so, having non-zero $\lambda$ values with $|\lambda |$ typically 10--40$^{\circ}$. HAT-P-57b \citep{2015AJ....150..197H} is also likely to be moderately misaligned with 27$^{\circ}$\,<\,$\lambda$\,<\,58$^{\circ}$. The KELT-9b system \citep{2017Natur.546..514G} is the most recently discovered example, with the hottest star known to host a transiting planet ($\sim$10,000\,K) and the planet itself on a near-polar orbit ($\lambda\,\sim$\,--85$^{\circ}$).  WASP-33b is the exception in the list of \citet{2016arXiv161202776C}, being highly retrograde with $\lambda\,=\,-109^{\circ}\,\pm\,1^{\circ}$ \citep{2010MNRAS.407..507C}, and the same is now seen in \wkb\, with $\lambda\,=\,-165^{\circ}\,\pm\,5^{\circ}$.

As has been widely discussed \citep[e.g.][]{2012ApJ...757...18A,2014sf2a.conf..217C,2015MNRAS.450.3306F,2016ApJ...818....5L}, stars hotter than 6100 K host hot Jupiters with a large range of obliquities, whereas cooler stars tend to have planets in aligned orbits (see, e.g., Fig.~8 of \citet{2016arXiv161202776C}).  The suggestion is that hotter stars are less effective at tidally damping a planet's obliquity, perhaps owing to their relatively small convective envelopes \citep[e.g.][]{2010ApJ...718L.145W}. The discovery of \wkb\ now reinforces this trend.

\subsection{Stellar pulsations}
\label{discuss:pulsations}
WASP-167/KELT-13 is one of a growing number of hot-Jupiter hosts that have shown non-radial pulsations.  The first was WASP-33b \citep{2010MNRAS.407..507C}, which shows $\delta$-Scuti pulsations with a dominant period near 21 cycles/day (86 mins) and an amplitude of several mmag \citep{2013A&A...553A..44K,2014A&A...561A..48V}.  Further, \citet{2011A&A...526L..10H} noted that one of the pulsation frequencies was very near 26 times the orbital frequency of the planet, which suggests that the planet might be exciting the pulsations.

HAT-P-2b is an eccentric massive planet (8\,M$_{\rm Jup}$, e\,$\sim$\,0.5) in a 5-d orbit.  \citet{2017arXiv170203797D} detect pulsations in {\it Spitzer\/} lightcurves of HAT-P-2b, at a level of 40\,ppm, much lower than in  WASP-33b, but at a similar timescale of $\sim$\,87\,mins.   Owing to the commensurability between the pulsation and orbital frequencies,  \citet{2017arXiv170203797D} again suggest that the planet is exciting the pulsations. 

A third example is WASP-118, which shows pulsations at a timescale of $\sim$\,1.9-d and an amplitude of $\sim$\,200 ppm in {\it K2\/}  observations \citep{2017MNRAS.469.1622M}. Another is HAT-P-56, a $\gamma$-Dor pulsator with a primary pulsation period of ~1.644\,$\pm$\,0.03-d, which were also seen in {\it K2\/} observations \citep{2015AJ....150...85H}.

It is worth noting that both planets WASP-33b and \wkb\ have retrograde orbits, whereas that of HAT-P-2b is highly eccentric, which may be relevant to the excitation of the pulsations. 

In \wk, judging from Fig.~\ref{fig:SW1304tomog1}, the pulsations appear to have a timescale of $\sim$\,4-hours, though with limited data we cannot be more precise. The pulsations in \wk\ have a longer timescale than in WASP-33 and HAT-P-2 and are near the borderline between $\delta$-Scuti and $\gamma$-Dor  behaviour, and so we are unsure which class to assign the star to.  

It may be that the pulsations are contributing to the scatter in the RV measurements seen in Fig.~\ref{fig:RV-maxK1}. Indeed,  \citet{2017arXiv170203797D} attribute radial-velocity scatter in HAT-P-2 to the pulsations.  \citet{2016MNRAS.463.3276H} also report excess RV scatter in WASP-118. 

We have looked for the pulsations in the  WASP and KELT photometry, but do not detect any signal, with an upper limit of 0.5\,mmags.  However, we caution that the WASP data are not particularly suitable for searching for periodicities of 4 to 8 hours.  This is comparable to the length of observation on each night, and is thus the timescale of greatest red noise in WASP data. For this reason WASP data are processed to reduce sinusoidal-like variations on such timescales \citep{2006MNRAS.373..799C}. Similar considerations apply to the KELT data, which in any case have lower photometric precision.   The higher-quality follow-up photometry was aimed at observing the transits, before we were aware of the presence of pulsations, and none of the observations are long enough to search for the pulsations.

It is also possible that the particular mode of pulsations can lead to scatter in the RV measurements but smaller photometric variations owing to  geometric cancellation.  Axisymmetric non-radial pulsations of order l$\geq$3 are subject to partial geometric cancellation: the greater the value of l, the larger the cancellation effect, and odd-numbered modes are near invisible in intensity measurements \citep{2010aste.book.....A}.

\section*{Acknowledgements}
WASP-South is hosted by the South African Astronomical Observatory and we are grateful for their ongoing support and assistance. Funding for WASP comes from consortium universities and from the UK's Science and Technology Facilities Council. The Euler Swiss telescope is supported by the Swiss National Science Foundation. TRAPPIST is funded by the Belgian Fund for Scientific Research (Fond National de la Recherche Scientifique, FNRS) under the grant FRFC 2.5.594.09.F, with the participation of the Swiss National Science Foundation (SNF). M. Gillon and E. Jehin are FNRS Research Associates. We acknowledge use of the ESO 3.6-m/HARPS under program 096.C-0762. L. Delrez acknowledges support from the Gruber Foundation Fellowship.

Work performed by P.A.C. was supported by NASA grant NNX13AI46G. D.J.S and B.S.G. were partially supported by NSF CAREER Grant AST-1056524. S.V.Jr. is supported by the National Science Foundation Graduate Research Fellowship under Grant No. DGE-1343012. Work performed by J.E.R. was supported by the Harvard Future Faculty Leaders Postdoctoral fellowship. 


\bibliographystyle{mnras}
\bibliography{litbiblio}


\bsp
\label{lastpage}
\end{document}